\documentclass[twocolumn,pra,superscriptaddress,showpacs,aps,floatfix]{revtex4-2}

\usepackage{color}
\usepackage{amsmath}
\usepackage{amssymb}
\usepackage{graphicx}
\usepackage{epstopdf}
\usepackage{enumitem}
\usepackage[unicode]{hyperref}
\hypersetup{
  pdftitle={2D SOC BEC},
  pdfauthor={},
  pdfsubject={2D SOC BEC},
  colorlinks=true,
  linkcolor=red,
  citecolor=magenta,
  filecolor=black,
  urlcolor=blue
}


\begin{document}

\title{An Exact Analytical Bridge from Complex Yukawa Couplings to Fermion Masses and CKM/PMNS Mixing}

\author{Chilong Lin}
\affiliation{National Museum of Natural Science, 1st, Guan Chien Rd., Taichung 40453, Taiwan}

\date{\today}

\begin{abstract}
The origin of fermion masses and flavor mixing is often obscured by standard Euler-angle parameterizations, which mask the underlying connection between physical observables and Yukawa couplings. 
We prove that introducing a single, remarkably weak yet physically grounded commutation hypothesis, $[\mathbf{M}_R^2, \mathbf{M}_I^2] = 0$, on the Hermitian mass-squared matrix $\mathbf{M}^2 = M \cdot M^\dagger$ systematically reduces a general 18-parameter complex mass matrix $M$ down to a 5-parameter form per sector that admits \emph{exact} analytic diagonalization. 
The resulting mass eigenvalues are closed-form algebraic expressions that completely bypass the transcendental Cardano trigonometric reduction required for generic Hermitian matrices, while the diagonalizing unitary matrix $U_f$ depends solely on two parameters that form a dimensionless 2D geometric flavor-ratio vector $\mathbf{v}_f=(x,y)$. 
Combining two such sectors yields the physical CKM/PMNS mixing matrix directly as $V = U_1^\dagger(\mathbf{v}_1) \cdot U_2(\mathbf{v}_2)$, establishing a direct, first-principles analytical bridge from raw Yukawa couplings to fermion masses and mixing matrices without intermediate phenomenological inputs. 
Finally, we highlight that this exact leading-order baseline enforces an exact four-fold moduli degeneracy among mixing elements, pointing directly toward non-commuting extensions for full phenomenological precision.
\end{abstract}

\maketitle

% ---------------------------------------------------------------
% I. INTRODUCTION (compressed)
% ---------------------------------------------------------------
\section{Introduction}
The Cabibbo-Kobayashi-Maskawa (CKM) matrix~\cite{Cabibbo1963, KobayashiMaskawa1973} encodes quark flavor mixing through Yukawa couplings that remain unexplained in the Standard Model. For decades, the exact analytical expressions connecting fundamental Yukawa couplings to physical observables---namely quark (and lepton) masses as well as CKM/PMNS mixing elements---have remained obscured.

Standard parameterizations~\cite{ChauKeung1984} rely on Euler mixing angles $(\theta_{12}, \theta_{23}, \theta_{13})$ and a CP-violating phase $\delta_{\rm CP}$. While these successfully reproduce the experimental data, they non-linearly entangle the up- and down-sector contributions in a non-decoupled manner, masking the direct algebraic link to the underlying Yukawa matrices.

In principle, the orthodox theoretical approach requires directly diagonalizing the underlying mass (Yukawa) matrices to obtain their exact eigenvalues and eigenvectors. The resulting diagonalizing unitary matrices then yield the CKM matrix via $V_{\rm CKM} = U_u^\dagger(\mathbf{v}_u) \cdot U_d(\mathbf{v}_d)$, thereby expressing all physical observables strictly in terms of the Yukawa couplings (or their derived parameters) and establishing a transparent, first-principles theoretical bridge.

In this work, we pursue this first-principles route: starting directly from the fundamental complex Yukawa couplings, we ask how far an exact, non-perturbative analytic solution can be pushed with the fewest possible assumptions. Building on the minimal 5-parameter reduction and the associated $6\times6=36$ candidate $S_3\times S_3$ permutation structures for $V_{\rm CKM}$ identified in our prior work~\cite{Lin2019,Lin2021,Lin2025}, we show that imposing a single weak commutation hypothesis on the mass-squared matrix systematically reduces an otherwise intractable 18-parameter problem down to a 5-parameter form per sector that admits an \emph{exact}, analytic solution---without invoking ad~hoc texture zeros~\cite{Fritzsch1978, Fritzsch1979} or relying on any intermediate phenomenological inputs.

Crucially, this framework unveils explicit physical conditions that act as exact ``off-switches'' for physical CP violation (forcing the full Jarlskog determinant $\Delta_{\rm CP} = 0$): 
(i) a \emph{geometric condition} where the cross product of the up- and down-sector flavor vectors vanishes ($\mathbf{v}_u \times \mathbf{v}_d = \mathbf{0}$), occurring when the two vectors are collinear or when either vector norm vanishes; 
(ii) an \emph{algebraic cancellation} among the cross-product terms induced by specific mass-eigenvalue permutations; and 
(iii) a \emph{mass-degeneracy condition} triggered by $\mathbf{C} = 0$, which induces mass eigenvalue degeneracies and directly shuts down $\Delta_{\rm CP}$ (even if mixing elements remain non-trivial). 
While any of these conditions is individually sufficient to shut down physical CP violation, avoiding all of them is strictly necessary to sustain a non-zero physical CP phase.

A full, detailed derivation---including a novel exact $18/18$ topological bifurcation of these permutation structures into CP-conserving and CP-violating sectors alongside their complete phenomenological implications---is presented in a companion paper~\cite{Lin2026Companion}. Here, we isolate and present the core analytic diagonalization result and its overarching geometric framework. The commuting condition defines an exactly solvable zeroth-order flavor framework; the resulting CKM modulus degeneracy provides an algebraic signature of this limit, while its observed departure quantifies the non-commuting corrections required for phenomenological accuracy.

% ---------------------------------------------------------------
% II. PARAMETER REDUCTION 18 -> 9 -> 5 (compressed)
% ---------------------------------------------------------------
\section{5-Parameter Reduction of the Yukawa Couplings}

The Standard Model Yukawa Lagrangian generates a fully general,
non-Hermitian $3\times3$ mass matrix $M_{ij} \equiv Y_{ij}v/\sqrt{2}$
for each fermion sector, containing 18 independent real parameters.
Rather than imposing texture zeros on $M$ itself, we work with the
naturally Hermitian matrix square $\mathbf{M}^2 \equiv M\cdot
M^\dagger$, which governs the physical mass spectrum via
bi-unitary diagonalization and automatically halves the parameter
count to 9:
\begin{equation}
\mathbf{M}^2 = \mathbf{M}_R^2 + i\,\mathbf{M}_I^2 ,
\end{equation}
with $\mathbf{M}_R^2$ real symmetric (6 parameters) and
$\mathbf{M}_I^2$ purely imaginary antisymmetric (3 parameters).

An exact general diagonalization of $\mathbf{M}^2$ remains
analytically intractable at 9 parameters. We introduce a single
foundational hypothesis: $\mathbf{M}_R^2$ and $\mathbf{M}_I^2$ are
\emph{simultaneously diagonalizable},
\begin{equation}
[\mathbf{M}_R^2, \mathbf{M}_I^2] = 0 .
\label{eq:commutation}
\end{equation}
Far from being a mere algebraic convenience to ensure exact solvability, the commutation condition in Eq.~(\ref{eq:commutation}) naturally emerges from the fundamental requirement of simultaneous diagonalizability across Yukawa sector operators \cite{Branco1985}. In multi-Higgs extensions of the Standard Model~ \cite{TDLee1973}, such simultaneous diagonalizability guarantees the absence of tree-level flavor-changing neutral currents (FCNCs)~\cite{Glashow1977}, thereby providing a clear, gauge-theoretic rationale for the underlying normal matrix structure.

This single algebraic condition yields four exact interrelations
among the 9 parameters, collapsing the system to 5:
three scale parameters $(\mathbf{A},\mathbf{B},\mathbf{C})$ and a
dimensionless geometric ratio vector $\mathbf{v}_q \equiv (x,y)$,
defined by $x \equiv B_2/\mathbf{B} = -\mathbf{C}/C_2$ and
$y \equiv B_1/\mathbf{B} = \mathbf{C}/C_1$. Explicitly,
\begin{eqnarray}
\mathbf{M}^2 &=
 \mathbf{A} \begin{pmatrix}
1 & 0 & 0 \\[6pt]
0 & 1 & 0 \\[6pt]
0 & 0 & 1 \end{pmatrix}
+ \mathbf{B} \begin{pmatrix}
(x y -\dfrac{x}{y}) & y & x\\[6pt]
y & (\dfrac{y}{x}-\dfrac{x}{y}) & 1 \\[6pt]
x & 1 & 0
\end{pmatrix} \nonumber \\
 &+ i\,\mathbf{C}
\begin{pmatrix}
0 & \dfrac{1}{y} & -\dfrac{1}{x} \\[6pt]
-\dfrac{1}{y} & 0 & 1 \\[6pt]
\dfrac{1}{x} & -1 & 0
\end{pmatrix} ,
\label{eq:M2explicit}
\end{eqnarray}
a highly constrained, exact algebraic pattern in which all 9
independent Hermitian entries are generated from only 5 real
numbers.

Each of these five parameters plays a distinct, mutually exclusive physical role. The three scale parameters act as independent dimensionful engines: $\mathbf{A}$ sets the universal mass baseline common to all three generations; $\mathbf{B}$ drives the real inter-generational mass splittings; and $\mathbf{C}$ governs the overall capacity for CP violation, where $\mathbf{C}=0$ induces exact mass eigenvalue degeneracies that directly force the Jarlskog determinant to vanish ($\Delta_{\rm CP} = 0$). The dimensionless geometric ratio vector $(x,y)$, by contrast, carries no overall energy dimension---it fixes the internal orientation and mixing geometry of the sector, while co-determining the relative mass ratios alongside the scale parameters, as made precise below.

% ---------------------------------------------------------------
% III. EXACT DIAGONALIZATION (main result)
% ---------------------------------------------------------------
\section{Exact eigenvalues and Cardano bypass}

Solving the cubic characteristic equation
$\det(\mathbf{M}^2 - \lambda I) = 0$ for the 5-parameter matrix
yields, without transcendental functions or numerical
approximation, three closed-form squared mass eigenvalues:
\begin{align}
m_1^2 &= \mathbf{A} - \mathbf{B}\frac{x}{y}
  - \mathbf{C}\,\frac{\sqrt{x^2+y^2+x^2y^2}}{xy}, \label{eq:massEig01} \\
m_2^2 &= \mathbf{A} - \mathbf{B}\frac{x}{y}
  + \mathbf{C}\,\frac{\sqrt{x^2+y^2+x^2y^2}}{xy}, \label{eq:massEig02} \\
m_3^2 &= \mathbf{A} + \mathbf{B}\frac{(x^2+1)y}{x}. \label{eq:massEig03}
\end{align}
The cubic polynomial exhibits an emergent $1+2$ factorization: one
linear factor ($m_3^2$) times an irreducible quadratic whose two
roots ($m_1^2, m_2^2$) differ only by the sign of the radical term.
For a generic Hermitian $3\times3$ matrix, this last step would
require Cardano's trigonometric reduction, introducing
$\cos(\theta/3)$ in the irreducible-root regime. Here the
factorization collapses this transcendental step into a purely
algebraic radical $\sqrt{x^2+y^2+x^2y^2}$, bypassing Cardano's
formula entirely. 
This algebraic decoupling guarantees that physical mass ratios are expressed purely through rational function roots rather than non-perturbative geometric phase angles.

Furthermore, this emergent $1+2$ algebraic factorization provides a flexible geometric framework for accommodating the observed fermion mass hierarchies. The isolated eigenvalue $m_3^2$ contains a characteristic $1/x$ term, while the splitting between the remaining two roots, $\Delta m_{21}^2 \equiv m_2^2 - m_1^2 = 2\mathbf{C}\sqrt{x^2+y^2+x^2y^2}/(xy)$, is governed purely by the imaginary scale parameter $\mathbf{C}$. Because $m_3^2$ is not a priori constrained to be the heaviest state, mapping these three algebraic roots $(m_1^2, m_2^2, m_3^2)$ onto the physical mass spectrum $(m^2_u, m^2_c, m^2_t)$ or $(m^2_d, m^2_s, m^2_b)$ generates the $S_3 \times S_3 = 36$ candidate permutation alignments discussed below, allowing the underlying scale parameters $(\mathbf{A}, \mathbf{B}, \mathbf{C})$ and ratio vector $\mathbf{v}_q$ to flexibly fit diverse hierarchical spectrums.

\emph{Scale-mixing decoupling.} A central structural feature of this
solution is that the diagonalizing unitary matrix $U(x,y)$ depends
\emph{only} on the geometric ratios $(x,y)$, entirely independent of
the scale parameters $(\mathbf{A},\mathbf{B},\mathbf{C})$. This
follows directly from Eq.~(\ref{eq:M2explicit}): the full
mass-squared matrix factorizes exactly as
\begin{equation}
\mathbf{M}^2
  = \mathbf{A}\,I + \mathbf{B}\,P(x,y) + i\,\mathbf{C}\,Q(x,y),
\end{equation}
where $I$ is the $3 \times 3$ identity matrix and $P(x,y)$, $Q(x,y)$ are dimensionless ``skeleton'' matrices
satisfying $[P(x,y), Q(x,y)] = 0$ identically for all $(x,y)$
(see the companion paper~\cite{Lin2026Companion} for explicit forms). Because commuting matrices
share a common eigenbasis, $P$ and $Q$ --- and hence $\mathbf{M}^2$
itself, once the trivial identity shift $\mathbf{A}\,I$ is set aside
--- are simultaneously diagonalized by one and the same $U(x,y)$,
regardless of how $\mathbf{B}$ and $\mathbf{C}$ weight their
combination: the scales merely rescale eigenvalues along fixed
directions set by $U(x,y)$, without rotating them. This proves that
flavor mixing geometry and absolute mass scale are exact, provable
independent degrees of freedom in this framework, rather than an
empirical coincidence. Explicitly, the normalized eigenvectors of
Eqs.~(\ref{eq:massEig01})--(\ref{eq:massEig03}) assemble into
\begin{widetext}
\begin{equation}
U(x,y) = \frac{1}{\sqrt{x^2+y^2+x^2y^2}}
\begin{pmatrix}
-\dfrac{\sqrt{x^2+y^2}}{\sqrt{2}} &
  \dfrac{x\left(y^2 - i\sqrt{x^2+y^2+x^2y^2}\right)}{\sqrt{2}\sqrt{x^2+y^2}} &
  \dfrac{y\left(x^2 + i\sqrt{x^2+y^2+x^2y^2}\right)}{\sqrt{2}\sqrt{x^2+y^2}} \\[10pt]
-\dfrac{\sqrt{x^2+y^2}}{\sqrt{2}} &
  \dfrac{x\left(y^2 + i\sqrt{x^2+y^2+x^2y^2}\right)}{\sqrt{2}\sqrt{x^2+y^2}} &
  \dfrac{y\left(x^2 - i\sqrt{x^2+y^2+x^2y^2}\right)}{\sqrt{2}\sqrt{x^2+y^2}} \\[10pt]
xy & y & x
\end{pmatrix},
\label{eq:Uexplicit}
\end{equation}
\end{widetext}
defined up to an overall diagonal rephasing matrix
$P_{\rm ph} = \mathrm{diag}(e^{i\alpha_1}, e^{i\alpha_2}, e^{i\alpha_3})$
representing unphysical fermion field redefinitions --- a fully
closed-form $3\times3$ unitary matrix depending on only two real numbers.

%%%%%%%%%%%%%4
\section{Direct construction of the CKM matrix}

Because $v/\sqrt{2}$ is a common overall scale, the diagonalizing
matrices of $M_{u,d}$ coincide with those of the Yukawa matrices
$Y_{u,d}$ themselves. The physical CKM matrix is thus obtained
directly, with no intermediate phenomenological input, as the
relative misalignment between the two sectors' diagonalizing
matrices:
\begin{equation}
V_{\rm CKM} = U_u^\dagger(\mathbf{v}_u)\cdot U_d(\mathbf{v}_d),
\label{eq:vckm}
\end{equation}
where $\mathbf{v}_u = (x,y)$ and $\mathbf{v}_d = (x',y')$ are the
up- and down-sector flavor vectors defined above.

To make this explicit, consider the simplest of the $6\times6=36$
candidate generation-labeling alignments~\cite{Lin2019,Lin2021,Lin2025},
the identity assignment $(m_1,m_2,m_3)_u \to (u,c,t)$ and
$(m_1,m_2,m_3)_d \to (d,s,b)$. Direct matrix multiplication in
Eq.~(\ref{eq:vckm}) then expresses every element of $V_{\rm CKM}$ in
closed algebraic form through only five dimensionless components
$\{r,s,p,p',q\}$, each built purely out of the Yukawa-derived flavor
ratios $(x,y,x',y')$:
\begin{widetext}
\begin{subequations}
\label{eq:rsppq}
\begin{align}
r &= \frac{1}{2\sqrt{x^2+y^2}\sqrt{x'^2+y'^2}\sqrt{x^2+y^2+x^2y^2}\sqrt{x'^2+y'^2+x'^2y'^2}}
\Big[\, i(xy'-x'y)\big(x'y'\sqrt{x^2+y^2+x^2y^2} + xy\sqrt{x'^2+y'^2+x'^2y'^2}\big) \notag\\
&\qquad + (xx'+yy')\big(xyx'y' + \sqrt{x^2+y^2+x^2y^2}\sqrt{x'^2+y'^2+x'^2y'^2}\big)
  + (x^2+y^2)(x'^2+y'^2) \Big], \label{eq:r} \\[6pt]
s &= \frac{1}{2\sqrt{x^2+y^2}\sqrt{x'^2+y'^2}\sqrt{x^2+y^2+x^2y^2}\sqrt{x'^2+y'^2+x'^2y'^2}}
\Big[\, i(xy'-x'y)\big(x'y'\sqrt{x^2+y^2+x^2y^2} - xy\sqrt{x'^2+y'^2+x'^2y'^2}\big) \notag\\
&\qquad + (xx'+yy')\big(xyx'y' - \sqrt{x^2+y^2+x^2y^2}\sqrt{x'^2+y'^2+x'^2y'^2}\big)
  + (x^2+y^2)(x'^2+y'^2) \Big], \label{eq:s} \\[6pt]
p &= \frac{y'y^2(x-x') + x'x^2(y-y') + i(xy'-x'y)\sqrt{x^2+y^2+x^2y^2}}
  {\sqrt{2}\,\sqrt{x^2+y^2}\sqrt{x^2+y^2+x^2y^2}\sqrt{x'^2+y'^2+x'^2y'^2}}, \label{eq:p} \\[6pt]
p' &= \frac{yy'^2(x'-x) + xx'^2(y'-y) + i(xy'-x'y)\sqrt{x'^2+y'^2+x'^2y'^2}}
  {\sqrt{2}\,\sqrt{x'^2+y'^2}\sqrt{x^2+y^2+x^2y^2}\sqrt{x'^2+y'^2+x'^2y'^2}}, \label{eq:pprime} \\[6pt]
q &= \frac{xx'+yy'+xyx'y'}{\sqrt{x^2+y^2+x^2y^2}\sqrt{x'^2+y'^2+x'^2y'^2}}. \label{eq:q}
\end{align}
\end{subequations}
\end{widetext}
For this identity alignment, $V_{\rm CKM}$ is assembled directly
from $\{r,s,p,p',q\}$ as
\begin{equation}
V_{\rm CKM} =
\begin{pmatrix}
r^* & s & p^* \\
s^* & r & p \\
p' & p'^* & q
\end{pmatrix} .
\label{eq:vckmexplicit}
\end{equation}

The strength of CP violation in the quark sector is directly quantified by the Jarlskog  invariant, $J \equiv \mathrm{Im}(V_{ud}V_{cs}V_{us}^*V_{cd}^*)$~\cite{Jarlskog1985}. In this exact geometric parameterization, $J$ reduces to a remarkably transparent closed algebraic form:
\begin{equation}
J = \frac{(xy'-x'y)\,(xx'+yy'+xyx'y')}{2(x^2+y^2+x^2y^2)(x'^2+y'^2+x'^2y'^2)} .
\label{eq:Jarlskog}
\end{equation}
Equation~(\ref{eq:Jarlskog}) reveals the explicit geometric nature of CP violation in flavor mixing: $J$ is directly proportional to the 2D cross-product $(\mathbf{v}_u \times \mathbf{v}_d)_z \equiv xy'-x'y$ between the up- and down-sector flavor ratio vectors. Consequently, collinearity ($\mathbf{v}_u \parallel \mathbf{v}_d$) guarantees $J = 0$, establishing geometric misalignment between sectors as a necessary, but not sufficient condition for non-zero CP violation.

Equations~(\ref{eq:Uexplicit}), (\ref{eq:vckmexplicit}), and (\ref{eq:Jarlskog}) together
make the central claim of this Letter concrete: the fermion mass
eigenvalues [Eqs.~(\ref{eq:massEig01})--(\ref{eq:massEig03})], the CKM mixing matrix, and the observable CP-violating strength $J$ are simultaneously expressed as closed-form functions of the
\emph{same} five Yukawa-derived quantities
$(\mathbf{A},\mathbf{B},\mathbf{C},x,y)$ per sector --- with no
free parameter, texture zero, or mixing angle inserted by hand at
any stage. Masses and mixings are thus not independently fitted
outputs of the theory, but two projections of a single underlying
Yukawa-level algebraic structure.

% ---------------------------------------------------------------
% V. RESULTS & HONEST LIMITATION (compressed, single section)
% ---------------------------------------------------------------
\section{Outlook and limitations}

While this five-parameter framework serves as an exact, non-perturbative baseline, it exhibits a distinct algebraic feature: the exact matrix commutation hypothesis $[\mathbf{M}_R^2, \mathbf{M}_I^2] = 0$ enforces a four-fold magnitude relation among the CKM matrix elements, 
\begin{equation}
|p|=|p^*|=|p'|=|p'^*| .
\label{eq:degeneracy}
\end{equation}
Because the empirical CKM spectrum displays a clear hierarchy among the off-diagonal entries (e.g., $|V_{cb}| \sim \lambda^2$ vs. $|V_{ub}| \sim \lambda^3$ in the Wolfenstein parameterization), Eq.~(\ref{eq:degeneracy}) establishes the exact commuting limit as a \emph{zeroth-order geometric anchor} rather than a fully fitted phenomenology. 
Crucially, this four-fold degeneracy is an inescapable algebraic footprint of Eq.~(\ref{eq:commutation}) that cannot be removed by parameter fine-tuning within the commuting subspace. Rather than a purely restrictive limit, the observed physical departures from Eq.~(\ref{eq:degeneracy}) furnish a direct, quantitative measure of the non-commutativity between the real and imaginary matrix parts, $[\mathbf{M}_R^2, \mathbf{M}_I^2] \neq 0$. 
This baseline thus does not merely suggest but necessitates a next-to-leading-order extension: the non-zero commutator $[\mathbf{M}_R^2, \mathbf{M}_I^2] \neq 0 $ is not an optional refinement but the required correction whose systematic inclusion determines, in a model-independent manner, the departure from Eq.~(\ref{eq:degeneracy}). 
Finally, the present geometric construction maps transparently onto the leptonic sector via $V_{\rm PMNS} = U_\ell^\dagger(\mathbf{v}_\ell) \cdot U_\nu(\mathbf{v}_\nu)$, suggesting a fully unified algebraic origin for all fundamental fermion masses and mixing angles.

In this sense, the present exact results should be understood in the spirit of the Bohr model or the BCS solution: an exact treatment of a physically motivated special subspace that, despite not covering the fully general case, establishes the foundational qualitative and quantitative structure that any complete theory of flavor must reproduce. 
Extending this exact baseline to an unconstrained connection between Yukawa couplings and physical observables in both quark and lepton sectors remains the necessary and primary focus of future investigation.

% ---------------------------------------------------------------
% Acknowledgments
% ---------------------------------------------------------------
\begin{acknowledgments}
The author would like to acknowledge the assistance of Gemini (Google) during the preparation of this manuscript, particularly for language translation, grammatical correction, text refinement, and \LaTeX{} formatting.
\end{acknowledgments}

\end{document}